\documentclass[10pt,b5paper]{euromech498}
\usepackage[cam,a4,center,axes]{crop}
\usepackage{bezier,graphics,psfrag}
\usepackage{lastpage}

\begin{document}

\title[]{Cracked rotor vibrations by multifractal analysis }

\author{Grzegorz Litak}
\address{%
 Applied Mechanics Department,
 Lublin University of Technology,
 Nadbystrzycka~36, \\
 20-618 Lublin,
 Poland}
 \email{g.litak@pollub.pl}


\author{Jerzy T. Sawicki}
\address{%
Cleveland State University, Department of Mechanical Engineering, \\ Cleveland,~OH~44115}
 \email{j.sawicki@csuohio.edu}

 
\commby{}

\begin{abstract}
Multifractal analysis has been used to diagnoze  cracked and healthy 
rotors. Is has been shown that the complexity and regularity criteria of the 
dynamical systems defined 
by the multiple scaling  of the time series can  indicate the damages of the 
rotating shaft. 
\end{abstract}

\maketitle

In many technical devices and machines possessing rotor, their
actual dynamic condition determines their proper and safe operation. Most rotation machines operate 
over the 
extended periods of time
in various temperature regions, and often they are subjected to  large loads. 
As a consequence, of working conditions, their components are exposed to potential structural
damage such as the shaft surface crack [1,2].

\begin{figure}[htb]
 \rotatebox{0}{\scalebox{0.32}{
 \includegraphics{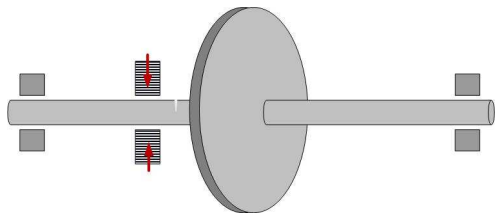}}}

\caption{ Schematic plot  of the cracked  rotor. In the experiment the shaft 
diameter was 15.875 mm and the shaft length was 0.659 m. The diameter of the active 
magnetic bearing
rotors and radial actuator was 47.625 mm. The disk has a
diameter of 127 mm and a thickness of 12.7 mm. The crack and magnetic actuator are depicted  on 
the right hand side of the disk. 
}
\end{figure} 

In the present paper we study
the  experimentally  determined response of the test rotor system focusing on the effect of
crack in a shaft.
The test rotor consists of the shaft
 supported two ball bearings and a single disk located midspan on a flexible shaft (Fig. 1). An
active magnetic actuator placed near the disk
produced an external harmonic force.
The crack had the width
of 0.94 mm  and the depth of 40\% of the shaft
diameter. The experimental time series of measured displacement,  
in the direction
of 45 deg from vertical one, 
for cracked and healthy 
rotors in the presence and are presented in Fig. 2.

\begin{figure}[htb]
\vspace{-1.3cm}

\centerline{
\hspace{1cm} \rotatebox{-90}{ 
\scalebox{0.48}{
 \includegraphics{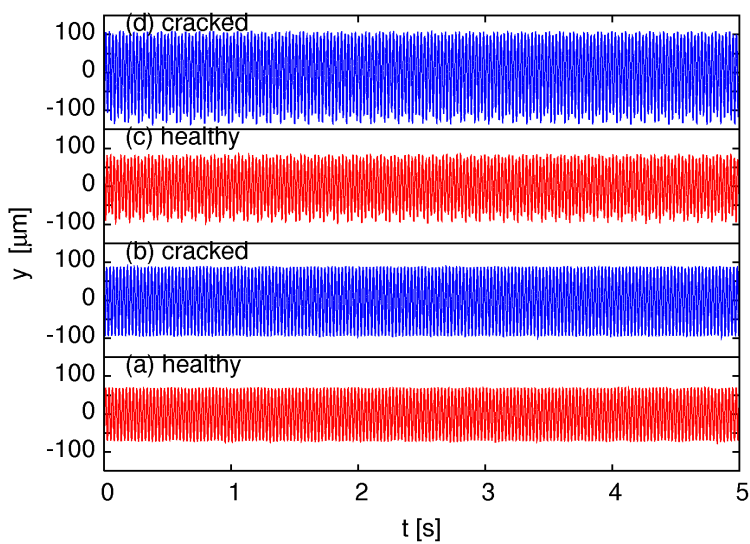}}}
}
\vspace{-0.6cm}

\caption{Experimental time series of the damaged and 
healthy machine for a spin velocity 2200 rpm (36.7 Hz):  (a,b) without external excitation;
(c,d) with magnetic actuator frequency
 3780 rpm (63 Hz).  
}
\vspace{-0.2cm}
\end{figure}

In Figs.  2a-b we plotted the results of unbalance  for healthy (Fig. 2a) and 
cracked (Fig. 2b) rotors, respectively. One can see some increase of the amplitude.   
More sophisticated changes can be visible in Figs. 2c-d where healthy (Fig. 2c) 
and
cracked (Fig. 2d) rotors have been subjected to 
the magnetic actuator applied harmonic
force having amplitude of 200 N p-p (peak-to-peak) and fixed frequency [1].

In further studies we propose to use a multifractal analysis [3] which appeared to 
be  a powerful tool to analyze the complexity of the
nonlinear systems. This technique has been widely used in 
biological systems [3,4] but recently has been applied in engineering systems, e.g., to examine 
seismic
sequences [5].
Following the multifractal procedure [3]  we performed  the 
Taylor expansion
of the time series in the small vicinity time instant $t_i$ we look for the 
exponent $h_i$ (usually non-integer), which limits
the error
between the examined function and Taylor expansion 
and determines the local singularity in the time series: 

$$
y(t)- a_0 - a_1(t-t_i) - a_2 (t-t_i)^2 - a_3 (t- t_i)^3 - ... \le a_h 
(t-t_i)^{h_i}.  
$$
Here  $a_i$ are locally determined Taylor expansion constants and $a_h$ is a 
coefficient  related to the exponent $h_i$.
The multifractal analysis of rotor vibrations is based on constructing a
singularity spectrum
$f(h)$ of all $h_i$ exponents 
providing a precise quantitative description of 
the  system behaviour [3,4]. Formally, $h$ defines 
the  H\"{o}lder exponent while the probability of its distribution $f(h)$ coincides 
with the Hausdorff dimension of a dynamical system.

The results of our calculations are presented in Figs. 3a and b for the rotor 
response without
external excitation and in the presence of magnetic actuators, respectively.  
The width of the spectrum $f(h)$, $\Delta h=h_{max}-h_{min}$  is defined as the complexity 
measure of the system response while the
$h_0$ which corresponds to the maximum of $f(h)$ indicates the regularity of vibrations.
The wider the range of possible fractal
exponents, the "richer" the process, in a structure.
Larger $h_0$ means more regular (or less stochastic) vibrations.

\begin{figure}[htb]
\vspace{-0.7cm}

 \rotatebox{-90}{\scalebox{0.23}{
 \includegraphics{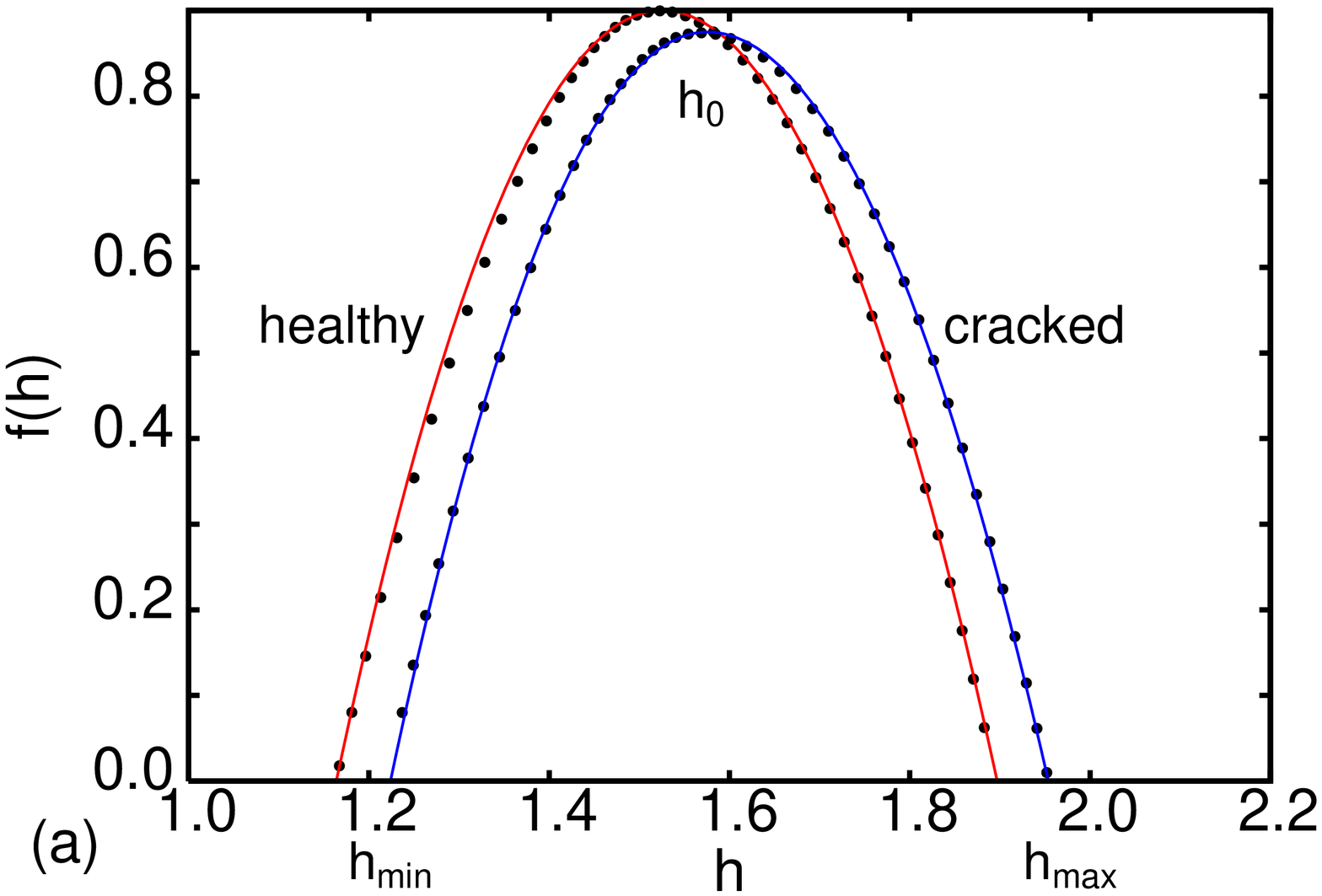}}} \hspace{-1cm}
 \rotatebox{-90}{\scalebox{0.23}{
 \includegraphics{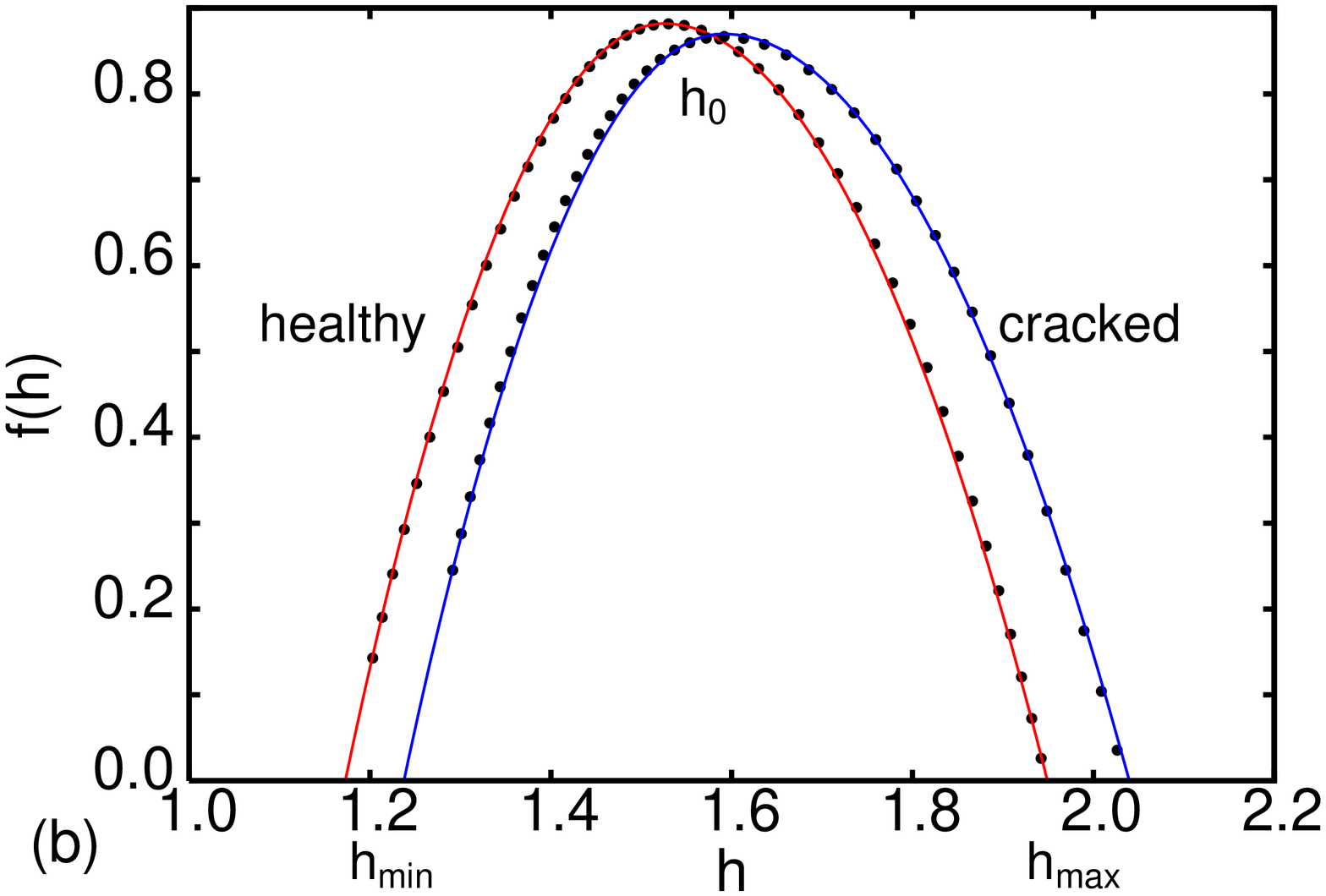}}}
\vspace{-0.2cm}

\caption{Spectra of the singularity exponents $f(h)$ for the system without 
any external excitation 
(a) and in the presence of magnetic actuators (b) (see the corresponding time series in Fig. 2).    
The width of corresponding spectra $\Delta h$ and a peak position $h_0$ for  
healthy and cracked rotors:
$\Delta h=$  0.729  \& 0.732, $h_0=$ 1.525 \& 1.575  (a);
$\Delta h=$   0.775  \&  0.803, $h_0=$ 1.525 \& 1.590  (b). 
\vspace{-0.2cm}
}
\end{figure}

In both plots we observe the shift of the spectrum to the right side (see the 
changes of $h_0$ position).
This is related to more periodic behaviour of the cracked system. This effect is 
a manifestation of some extra coupling between vibration modes created by a defect 
in the rotor systems. 
Note also that in the rotor system without external excitation Fig. 3a. the 
changes in the multifractal results in  terms of $\Delta h$ are negligible 
(increase by  0.4 \%) for the healthy and cracked rotors. Interestingly this  
difference of $\Delta h$,
in case of the excited rotor, is noticeable (about one order higher increase - 3.6 
\%).

The above findings let us to conclude that by monitoring the changes in the multifractal  spectra 
of a
rotor we could identify crack (and other faults) in a
rotor at an early stage in their development. 
The presented
results show that the use of an
multifractal analysis  can identify  the complex response of the system. The role 
of the magnetic actuator is crucial to increase visibility of the crack--induced 
effect.

In summary, we conclude that the presented approach has some advantages, enabling to 
quantify 
the effect of crack using the measure of complexity. The alternative way is to 
analyze the corresponding Fourier spectrum [1]. However
to produce a robust condition
monitoring technique more tests are necessary. Especially one could perform different 
excitation type and procedure
by using magnetic actuators. 
In our calculations we used the software provided by physionet [3].

\end{document}